# A dynamic model to study the potential TB infections and assessment of control strategies in China

Chuanqing Xu[1*], Kedeng Cheng[1], Songbai Guo[1], Dehui Yuan[2], Xiaoyu Zhao[3]

1 School of Science, Beijing University of Civil Engineering and Architecture, Beijing, China

2 Mathematics department, Hanshan Normal University, Chaozhou, China

3 School of Mathematics, Shandong University, Jinan, Shandong, 250100, China

**Abstract**：China is one of the countries with a high burden of tuberculosis, and although the number of new cases of tuberculosis has been decreasing year by year, the number of new infections per year has remained high and the diagnosis rate of tuberculosis-infected patients has remained low. Based on the analysis of TB infection data, we develop a model of TB transmission dynamics that include potentially infected individuals and BCG vaccination, fit the model parameters to the data on new TB cases, calculate the basic reproduction number $\mathcal{R}_v$= 0.4442. A parametric sensitivity analysis of $\mathcal{R}_v$ is performed, and we obtained the correlation coefficients of BCG vaccination rate and effectiveness rate with $\mathcal{R}_v$ as -0.810, -0.825. According to the model, we estimate that there are 614,186 (95% CI [562,631,665,741]) potentially infected TB cases in China, accounting for about 39.5% of the total number of TB cases. We assess the feasibility of achieving the goals of the WHO strategy to end tuberculosis in China and find that reducing the number of new cases by 90 per cent by 2035 is very difficult with the current tuberculosis control measures. However, with an effective combination of control measures such as increased detection of potentially infected persons, improved drug treatment, and reduction of overall exposure to tuberculosis patients, it is feasible to reach the WHO strategic goal of ending tuberculosis by 2035.

**Key words**：tuberculosis；Basic reproduction number；potential infection；sensitivity analysis；feasibility assessment

## 1.Introduction

Tuberculosis (TB) is an ancient disease with a global distribution and is the leading cause of death from bacterial infectious diseases. Mycobacterium tuberculosis, the causative agent of tuberculosis, can invade all organs of the body, but it is most common to cause tuberculosis in the lungs. Tuberculosis is mainly transmitted through the respiratory tract, and the source of infection is a person infected with tuberculosis in the infectious stage. Adolescents between the ages of 15 and 35 have a high incidence of tuberculosis, and the incubation period is typically 4-8 weeks. In the 19th century, tuberculosis became so prevalent in Europe and elsewhere that it killed one in seven people and became known as the "great white plague" [1]. The disease is still not fully under control and is still widespread in some countries around the globe. Between 1993 and 1996, the number of tuberculosis cases increased by 13 per cent globally, and the number of deaths from tuberculosis was greater than that from AIDS and malaria combined. At the end of 1995, in order to further promote global tuberculosis prevention and control publicity activities, the World Health Organization (WHO) established 24 March as "World Tuberculosis Day" [2]. About 80% of the world's new tuberculosis cases occur in 22 high-burden countries, with India and China having the highest number of cases, accounting for 26% and 12%, respectively, of the global total [3]. TB remains the leading cause of illness and death in most countries with high TB prevalence. WHO proposed a post-2015 global End TB strategy goal in 2014: to reduce the incidence of TB by 50 percent by 2025 (compared to 2015) and to reduce new cases by 90 percent by 2035.

---

* Corresponding author: xuchuanqing@bucea.edu.cn

Mathematical models have become a powerful tool for analyzing epidemiological characteristics [4]. Many scholars developed mathematical models reflecting the characteristics of TB based on its transmission mechanism, principles of biology, seasonal characteristic and social influences. In 1962, Waaler [5] developed the first model of tuberculosis transmission kinetics, which laid a solid foundation for subsequent research on tuberculosis. Xu et al [6] developed a model of tuberculosis transmission that included different age groups and explored the role of age structure in the process of tuberculosis transmission. Zain et al [7] analyzed and compared several models of tuberculosis. It has been shown that an increase in the treatment rate for active TB from 50% to 60% would significantly reduce the incidence of TB. Many studies have considered the effects of resistant cases [8], time lag [9] and age structure [10]. Tuberculosis (TB) has a wide range of modes of transmission and mild initial features, which make it easy to be overlooked and the spread of the disease somewhat insidious. In addition, the treatment of TB is long and highly susceptible to the development of drug resistance, and the success rate of treatment for drug-resistant TB is low. These characteristics mean that a significant number of potentially tuberculosis-infected people in China remain undiagnosed. 2023 Xu et al [11] developed a model of hepatitis B transmission dynamics that included potentially infected individuals and vaccination and assessed the number of potentially infected individuals with hepatitis B in China to be approximately 450,000. However, there are no studies of potential TB infections based on the transmission process.

We develop a model of tuberculosis transmission incorporating potential infected persons and vaccination, estimate the number of potentially infected persons with tuberculosis in China, explore the impact of different control measures on the number of new cases of tuberculosis, and assess the control strategies needed to meet the World Health Organization's targets.

The main research of this paper is as follows: in section 2, we analyze the collected data on tuberculosis, develop a model of tuberculosis transmission dynamics including potentially infected compartments and perform a qualitative analysis. In Section 3, we fit the main parameters of the model, calculate the value of the basic reproduction number $\mathcal{R}_v$, perform a sensitivity analysis, and make a feasibility assessment for achieving the strategic objectives of WHO. Section 4 is a discussion section.

## 2. Materials and methods

### 2.1 Data analysis

Currently, tuberculosis remains a major public health problem globally, plaguing several countries around the world. In 2023, the World Health Organization released the latest edition of the Global Tuberculosis Report [12], demonstrating the distribution of new cases of tuberculosis disease in 2022, and the results are shown in Figure 1. It shows that in 2022, majority of TB patients are distributed in WHO Southeast Asia (46%), Africa (23%) and the Western Pacific (18%), while the number of patients in the Eastern Mediterranean (8.1%), the Americas (3.1%), and Europe (2.2%) is low. Thirty high TB-burden countries account for 87% of all estimated cases globally, with eight countries accounting for more than two-thirds of the global total: India (27%), Indonesia (10%), China (7.1%), the Philippines (7.0%), Pakistan (5.7%), Nigeria (4.5%), Bangladesh (3.6%), and the Democratic Republic of the Congo (3.0%) [12]. This data indicates that China is still at a high level globally and faces a more serious tuberculosis epidemic.

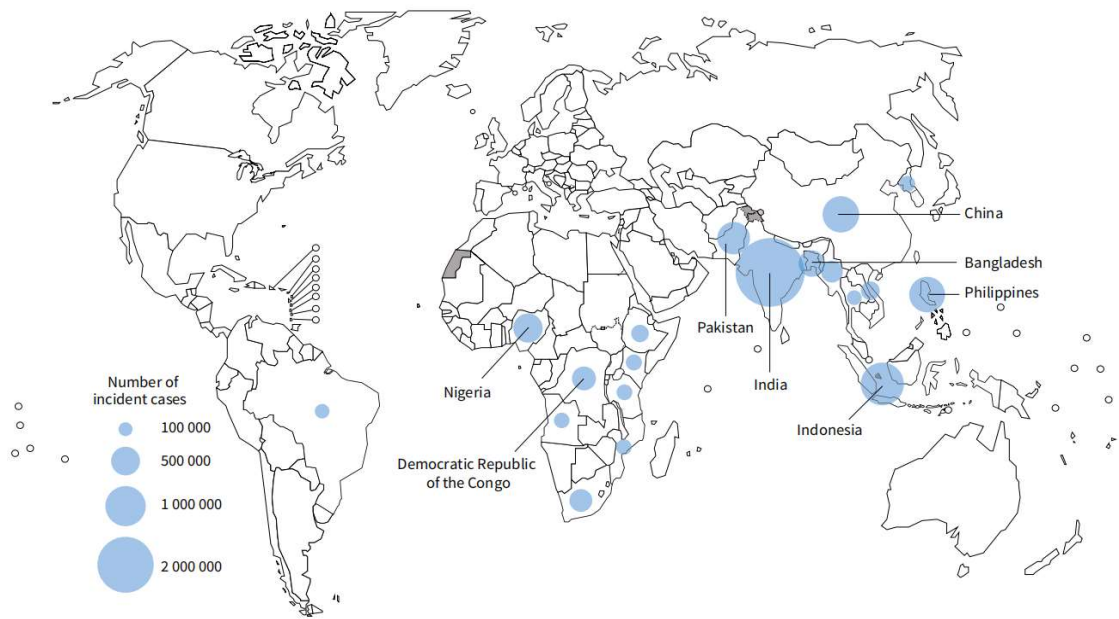

Figure 1 Estimated number of incident TB cases in 2022, for countries with at least 100,000 incident cases. (Image from Global TB Report 2023 [12])

We collected data related to tuberculosis from 2004-2021 from the website of China Public Health Science Data Center [13] and analyzed the data from different perspectives, and the results are shown in Figure 2. Figure 2(A) shows the increase in the number of new cases of tuberculosis over time, 2004-2021, and Figure 2(B) shows the trend in the number of deaths from tuberculosis over time, 2004-2021. The green area in the figure indicates that the tuberculosis prevention and treatment measures implemented in China during this period have been very effective in controlling the increase in the number of tuberculosis cases. It shows that the number of new cases of tuberculosis shows a gradual decline, with a maximum value of 1.25 million cases in 2005 and a minimum value of 630,000 cases in 2021. The number of deaths from tuberculosis has varied over the decade, with a maximum of 3,783 cases in 2009 and a minimum of 1,435 cases in 2004. There is a rapid decline in the number of incidence and deaths from TB in 2020, and we speculate that the decrease in data over that period is related to the prevention and control strategy of the Covid-19 epidemic pandemic.

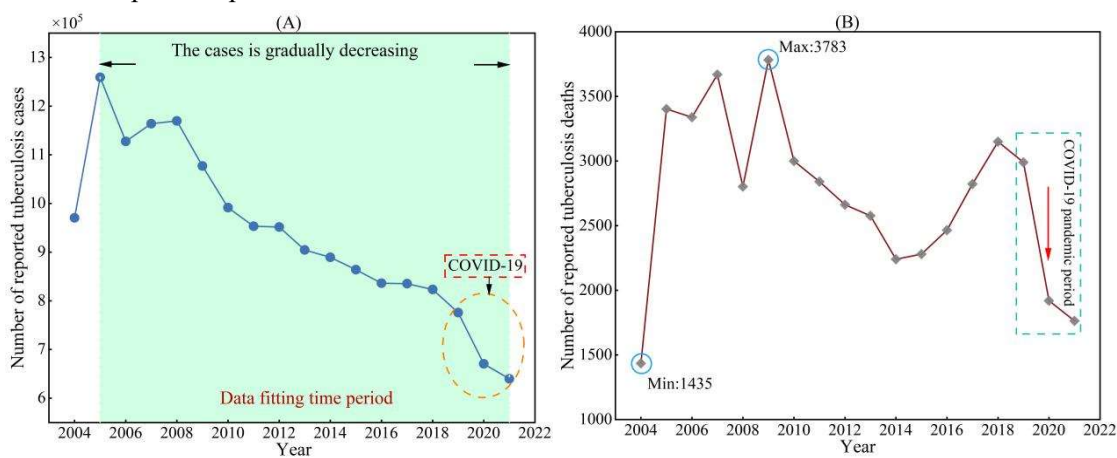

Figure 2. Tuberculosis incidence and deaths in China, 2004-2021. (A) Number of cases of tuberculosis, 2004-2021. (B) Number of deaths from tuberculosis, 2004-2021.

We collected data on TB infections in different regions of China from 2005-2019 and mapped them spatially geographically, and the results are shown in Figure 3. Geographically, the spatial distribution of the number of tuberculosis cases is uneven, with the number of cases decreasing more rapidly in the southern and eastern regions than in the northern and western regions. The regions with the highest average number of cases are Guangdong, Sichuan and Henan, with an annual average of 73,000 cases; the regions with the lowest average number of cases are Tibet, Ningxia and Tianjin, with an annual average of 3,500 cases. It is noteworthy that although the annual number of cases in Tibet and Qinghai is relatively small, the number of cases has increased over time. We believe that the uneven spatial distribution of the number of tuberculosis cases is related to the imbalance in the economic level of each region.

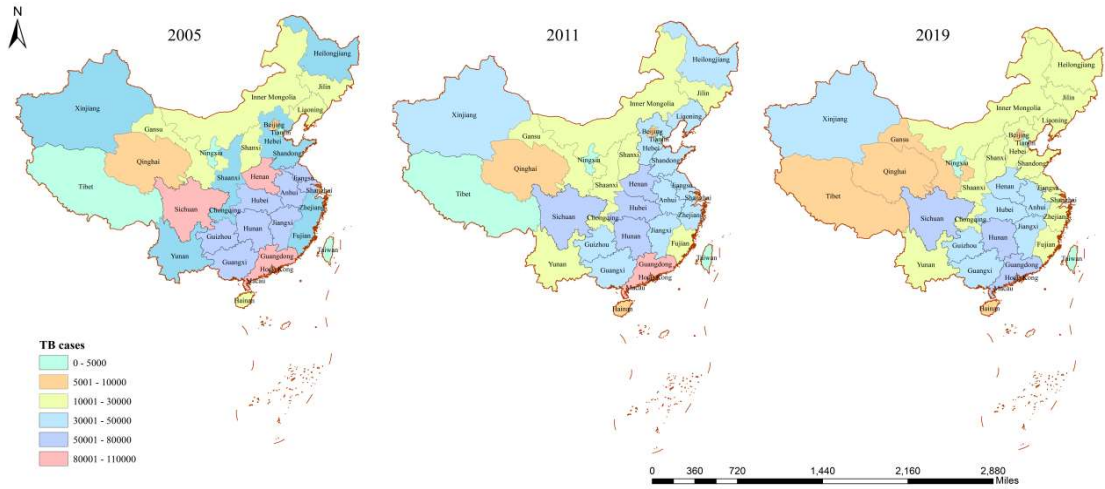

Figure 3. Distribution of tuberculosis infections by region in China, 2005-2019. (data from Hong Kong, Macau and Taiwan are not given).

**2.2 Model establishing**

According to a report on tuberculosis by the Chinese Center for Disease Control and Prevention (CDC), the number of people potentially infected with Mycobacterium tuberculosis in China is more than 200 million, and only 59% of tuberculosis patients are diagnosed [14]. Although more than 600,000 new TB infections are reported each year in China, it is likely that a larger number of infections go undetected. The main objective of this paper is to develop a dynamic model including potential TB infection based on the transmission mechanism of TB and to assess the number of potential TB infected individuals. The scheme of the model is shown in Figure 4. The model divides the population into five compartments, where S denotes susceptible, E denotes latent, $I_1$ denotes potentially TB-infected or undetected TB-infected, $I_2$ denotes detected TB-infected, and R denotes recovered, and the compartments contained in the blue dashed box are infectious, while the other compartments are not infectious. Since the people in the $I_2$ compartment are people who have reported infections and who consciously control their behavior, we assume in our model that the $I_2$ population is less contagious.

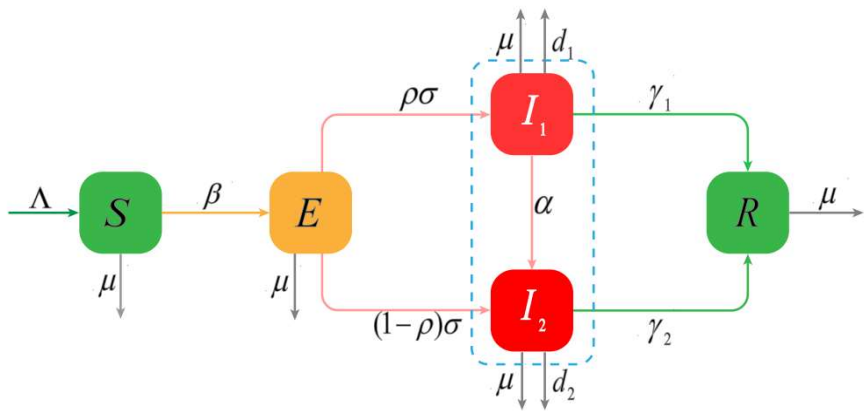

Figure 4. The scheme of the TB model

The corresponding propagation dynamics equation is constructed as follows:

$$\begin{cases} \frac{dS}{dt} = (1-\theta\omega)\Lambda - \beta S(I_1 + \varepsilon I_2) - \mu S, \\ \frac{dE}{dt} = \beta S(I_1 + \varepsilon I_2) - (\mu + \sigma)E, \\ \frac{dI_1}{dt} = \rho\sigma E - (\mu + \gamma_1 + \alpha + d_1)I_1, \\ \frac{dI_2}{dt} = (1-\rho)\sigma E + \alpha I_1 - (\mu + \gamma_2 + d_2)I_2, \\ \frac{dR}{dt} = \gamma_1 I_1 + \gamma_2 I_2 - \mu R. \end{cases} \quad (1)$$

$\Lambda$ is the annual number of births. $\theta$ denotes BCG vaccination rate at birth. $\omega$ denotes BCG efficiency. $\mu$ is natural mortality rate. $\beta$ is the base transmission rate. $\varepsilon$ is the relative transmission rate. $\sigma$ is the activation rate in latent patients. $\rho$ is the proportion of tuberculosis-infected who are undetected. $\alpha$ is the conversion rate of undetected compartment. $\gamma_1$、$\gamma_2$ are recovery rates for undetected and detected infected respectively. $d_1$、$d_2$ are mortality rates due to illness, respectively.

**2.3 Model analysis**

2.3.1 Calculation of disease-free equilibrium and basic reproduction number

Obviously, there is a disease-free equilibrium $P_0 = (S_0, 0,0,0,0)$ in system (1), where $S_0 = \frac{(1-\theta\omega)\Lambda}{\mu}$. In epidemiological studies, the basic reproduction number (denoted as $\mathcal{R}_v$) indicates the average number of infections in an infected person during the period of infection[15], and is one of the most important indicators to assess the risk of an infectious disease. We use the next generation matrix method[16] to calculate the basic reproduction number of system (1). The Jacobian matrices $F$ and $V$ at the disease-free equilibrium are obtained from system (1) as

$$F = \begin{pmatrix} 0 & \beta S_0 & \varepsilon\beta S_0 \\ 0 & 0 & 0 \\ 0 & 0 & 0 \end{pmatrix}, V = \begin{pmatrix} \sigma + \mu & 0 & 0 \\ -\rho\sigma & \gamma_1 + \mu + \alpha + d_1 & 0 \\ -(1-\rho)\sigma & -\alpha & \gamma_2 + \mu + d_2 \end{pmatrix}.$$

The basic reproduction number is the spectral radius of the $FV^{-1}$[15], therefore we get

$$\mathcal{R}_v = \frac{\beta S_0}{\sigma + \mu}\left(\frac{\rho\sigma}{\gamma_1 + \mu + \alpha + d_1} + \frac{(1-\rho)\sigma\varepsilon}{\gamma_2 + \mu + d_2} + \frac{\rho\sigma\alpha\varepsilon}{(\gamma_1 + \mu + \alpha + d_1)(\gamma_2 + \mu + d_2)}\right).$$

**2.3.2 Existence of the endemic equilibrium**

**Theorem 1**. When $\mathcal{R}_v > 1$, the system (1) has a unique positive equilibrium $P^* = (S^*, E^*, I_1^*, I_2^*, R^*)$.

**Proof.** Using the equilibrium equation of system (1) at the endemic equilibrium, we can get

$$S^* = \frac{S_0}{\mathcal{R}_v}, E^* = \frac{(1-\theta\omega)\Lambda}{\mu + \sigma}\left(1 - \frac{1}{\mathcal{R}_v}\right), I_1^* = \frac{\rho\sigma}{\gamma_1 + \mu + \alpha + d_1}E^*,$$

$$I_2^* = \left(\frac{\rho\sigma\alpha}{(\gamma_2 + \mu + d_2)(\gamma_1 + \mu + \alpha + d_1)} + \frac{(1-\rho)\sigma}{\gamma_2 + \mu + d_2}\right)E^*,$$

$$R^* = \frac{E^*}{\mu}\left[\frac{\rho\sigma\gamma_1}{\gamma_1 + \mu + \alpha + d_1} + \frac{\rho\sigma\alpha\gamma_2}{(\gamma_2 + \mu + d_2)(\gamma_1 + \mu + \alpha + d_1)} + \frac{(1-\rho)\sigma\gamma_2}{\gamma_2 + \mu + d_2}\right]. \quad (2)$$

Therefore, $E^*$ has a unique positive root if $\mathcal{R}_v > 1$. According to equation (2), we get that there is a unique positive equilibrium $P^*$ of system (1) at this time.

### 2.3.3 Stability analysis

It is not difficult to find that the set $\Gamma = \{\phi \in \mathbb{R}_+^5 : \sum_{i=1}^{5}\phi_i \leq \frac{(1-\theta\omega)\Lambda}{\mu}\}$ is positively invariant for model (1). And we have the following result.

**Theorem 2**. When $\mathcal{R}_v < 1$, the disease-free equilibrium $P_0$ of system (1) is locally asymptotically stable.

**Proof.** Let the Jacobian matrix of system (1) at the disease-free equilibrium $P_0$ be $M_1$. When $\mathcal{R}_v < 1$, assume that there is $\lambda_0$ satisfying $Re(\lambda_0) \geq 0$, such that $det(\lambda_0 E - M_1) = 0$. Expanding the characteristic equation, we can obtain

$$(\lambda_0 + \mu)^2 (\lambda_0 + \gamma_1 + \mu + \alpha + d_1)(\lambda_0 + \gamma_2 + \mu + d_2) \cdot \{\lambda_0 + (\sigma + \mu)[1 - F(\lambda_0)]\} = 0, \quad (3)$$

Here $F(\lambda_0) = \frac{\beta S_0}{\sigma + \mu}\left[\frac{\rho\sigma}{\lambda_0 + \gamma_1 + \mu + \alpha + d_1} + \frac{\rho\sigma\alpha\varepsilon}{(\lambda_0 + \gamma_1 + \mu + \alpha + d_1)(\lambda_0 + \gamma_2 + \mu + d_2)} + \frac{(1-\rho)\sigma\varepsilon}{\lambda_0 + \gamma_2 + \mu + d_2}\right].$

Since $Re(\lambda_0) \geq 0$ is assumed, the characteristic equation (3) can be simplified to

$$\lambda_0 + (\sigma + \mu)[1 - F(\lambda_0)] = 0.$$

The necessary condition for the above formula to be established is

$$Re[F(\lambda_0)] \geq 1. \quad (4)$$

Furthermore, since we assume $Re(\lambda_0) \geq 0$, we can obtain

$$|F(\lambda_0)| = \left|\frac{\beta S_0}{\sigma + \mu}\left[\frac{\rho\sigma}{\lambda_0 + \gamma_1 + \mu + \alpha + d_1} + \frac{\rho\sigma\alpha\varepsilon}{(\lambda_0 + \gamma_1 + \mu + \alpha + d_1)(\lambda_0 + \gamma_2 + \mu + d_2)}\right.\right.$$
$$\left.\left. + \frac{(1-\rho)\sigma\varepsilon}{\lambda_0 + \gamma_2 + \mu + d_2}\right]\right|$$
$$\leq \frac{\beta S_0}{\sigma + \mu}\left[\left|\frac{\rho\sigma}{\lambda_0 + \gamma_1 + \mu + \alpha + d_1}\right| + \left|\frac{\rho\sigma\alpha\varepsilon}{(\lambda_0 + \gamma_1 + \mu + \alpha + d_1)(\lambda_0 + \gamma_2 + \mu + d_2)}\right| + \left|\frac{(1-\rho)\sigma\varepsilon}{\lambda_0 + \gamma_2 + \mu + d_2}\right|\right]$$
$$\leq \frac{\beta S_0}{\sigma + \mu}\left[\left|\frac{\rho\sigma}{\gamma_1 + \mu + \alpha + d_1}\right| + \left|\frac{\rho\sigma\alpha\varepsilon}{(\gamma_1 + \mu + \alpha + d_1)(\gamma_2 + \mu + d_2)}\right| + \left|\frac{(1-\rho)\sigma\varepsilon}{\gamma_2 + \mu + d_2}\right|\right]$$
$$= \frac{\beta S_0}{\sigma + \mu}\left[\frac{\rho\sigma}{\gamma_1 + \mu + \alpha + d_1} + \frac{\rho\sigma\alpha\varepsilon}{(\gamma_1 + \mu + \alpha + d_1)(\gamma_2 + \mu + d_2)} + \frac{(1-\rho)\sigma\varepsilon}{\gamma_2 + \mu + d_2}\right]$$
$$= \mathcal{R}_v \quad (5)$$

Therefore, in the case of $\mathcal{R}_v < 1$, it can be known from inequality (5) that

$$Re[F(\lambda_0)] < 1. \quad (6)$$

It can be seen that there is a contradiction between inequalities (4) and (6), and the initial assumption about $\lambda_0$ is invalid, that is, all the characteristic roots of the characteristic equation $det(\lambda_0 E - M_1) = 0$ have negative real parts[17]. That means, if $\mathcal{R}_v < 1$, the disease-free equilibrium $P_0$ of system (1) is locally asymptotically stable[18].

**Theorem 3.** When $\mathcal{R}_v < 1$, the disease-free equilibrium $P_0$ of system (1) is globally asymptotically stable.

Define function

$$h(x) = x - 1 - \ln x, x > 0.$$

For any $x > 0$ there is $h(x) \geq 0$, and $h(x) = 0$ if and only if $x = 1$.

Proof. Let $U(t) = (S(t), E(t), I_1(t), I_2(t), R(t))$ be the solution of system (1) through any $\phi := (\phi_1, \phi_2, \phi_3, \phi_4, \phi_5) \in \Gamma$. Clearly, $U(t)$ is bounded and $S(t) > 0$ holds for any $t \geq 0$. We define a function $V$ on $\Omega = \{\phi \in \mathbb{R}_+^5 : \phi_1 > 0\}$ as follows[17],

$$V(\phi) = S_0 h\left(\frac{\phi_1}{S_0}\right) + R_v \phi_2 + \frac{(\gamma_2 + \mu + d_2 + \alpha\varepsilon)\beta S_0}{(\gamma_1 + \mu + \alpha + d_1)(\gamma_2 + \mu + d_2)}\phi_3 + \frac{\varepsilon\beta S_0}{\gamma_2 + \mu + d_2}\phi_4$$

The derivative of $V$ with respect to $t(t > 0)$ along $U(t)$ is

$$\frac{dV(U(t))}{dt} = -\frac{\mu}{S(t)}[S(t) - S_0]^2 + (R_v - 1)\beta S(t)\big(I_1(t) + \varepsilon I_2(t)\big).$$

Hence when $\mathcal{R}_v < 1$, there is $\frac{dV(t)}{dt} \leq 0$, which means that $V$ is a Lyapunov function on $\{U(t): t \geq 1\} \in \Omega$. And $\frac{dV(t)}{dt} = 0$ if and only if $U(t) = P_0$. Therefore, the largest invariant set of $\left\{U(t)\Big|\frac{dV(t)}{dt} = 0, U(t) \in \Omega\right\}$ is $\{P_0\}$. According to the LaSalle's Invariance Principle[19], when $\mathcal{R}_v < 1$, the disease-free equilibrium $P_0$ of system (1) is globally asymptotically stable.

**Theorem 4.** If $\mathcal{R}_v > 1$, the disease-free equilibrium $P_0$ of system (1) is unstable.

**Proof.** The characteristic polynomial of the Jacobian matrix at the disease-free equilibrium point $P_0$ is $f(\lambda)$

$$f(\lambda) = (\lambda + \mu)^2(\lambda^3 + a_1\lambda^2 + a_2\lambda + a_3) \tag{7}$$

Where $a_1 = b_1 + b_2 + b_3, a_2 = b_1 b_2 + b_1 b_3 + b_2 b_3 - \rho\sigma\beta S_0 - (1-\rho)\sigma\varepsilon\beta S_0, b_1 = \mu + \sigma, b_2 = \mu + \gamma_1 + \alpha + d_1, b_3 = \gamma_2 + \mu + d_2, a_3 = (\sigma + \mu)(\gamma_1 + \mu + \alpha + d_1)(\gamma_2 + \mu + d_2)(1 - R_v)$

Therefore, when $\mathcal{R}_v > 1$, there is $a_3 < 0$. The equation $f(\lambda) = 0$ has at least a positive root. As a result, the disease-free equilibrium $P_0$ of system (1) is unstable[18].

**Theorem 5.** When $\mathcal{R}_v > 1$, the endemic equilibrium $P^*$ of system (1) is locally asymptotically stable.

**Proof.** Let the Jacobian matrix of system (1) at the endemic equilibrium $P^*$ be $M_2$. When $\mathcal{R}_v > 1$, assume that there is $\lambda_2$ satisfying $Re(\lambda_2) \geq 0$, such that $det(\lambda_2 E - M_2) = 0$. Expanding the characteristic equation, we can obtain

$$(\lambda_2 + \mu)(\lambda_2 + \gamma_1 + \mu + \alpha + d_1)(\lambda_2 + \gamma_2 + \mu + d_2)$$
$$\cdot \left\{(\lambda_2 + \mu)\left[\lambda_2 + (\sigma + \mu)\left(1 - \frac{S^*}{S_0}G(\lambda_2)\right)\right] + \beta(I_1^* + \varepsilon I_2^*)(\lambda_2 + \sigma + \mu)\right\} = 0, \tag{8}$$

where,

$$G(\lambda_2) = \frac{\beta S_0}{\sigma + \mu}\left[\frac{\rho\sigma}{\lambda_2 + \gamma_1 + \mu + \alpha + d_1} + \frac{\rho\sigma\alpha\varepsilon}{(\lambda_2 + \gamma_1 + \mu + \alpha + d_1)(\lambda_2 + \gamma_2 + \mu + d_2)} + \frac{(1-\rho)\sigma\varepsilon}{\lambda_2 + \gamma_2 + \mu + d_2}\right].$$

The characteristic equation (8) can be simplified to

$$(\lambda_2 + \mu)(\lambda_2 + \sigma + \mu) + \beta(I_1^* + \varepsilon I_2^*)(\lambda_2 + \sigma + \mu) = \frac{S^*(\sigma + \mu)}{S_0}(\lambda_2 + \mu)G(\lambda_2).$$

Using equation (2), the above equation can be transformed into

$$(\lambda_2 + \mu)(\lambda_2 + \sigma + \mu) + \frac{R_v E^*}{S_0}(\sigma + \mu)(\lambda_2 + \sigma + \mu) = \frac{S^*(\sigma + \mu)}{S_0}(\lambda_2 + \mu)G(\lambda_2).$$

The above formula can be obtained by simplification as

$$1 + \frac{R_v E^*(\sigma + \mu)}{S_0} \cdot \frac{1}{\lambda_2 + \mu} = \frac{S^*(\sigma + \mu)}{S_0(\lambda_2 + \sigma + \mu)}G(\lambda_2). \tag{9}$$

Using equation (2) to replace the parameters of the above equation, we can obtain

$$\mathcal{R}_v = \frac{S_0}{S^*}.$$

Similar to the proof of inequality (5), taking the norm on the right side of the equal sign of equation (9) can get

$$\left|\frac{S^*(\sigma+\mu)}{S_0(\lambda_2+\sigma+\mu)}G(\lambda_2)\right| \leq \left|\frac{S^*(\sigma+\mu)}{S_0(\lambda_2+\sigma+\mu)}\mathcal{R}_v\right| = \left|\frac{\sigma+\mu}{\lambda_2+\sigma+\mu}\right| \leq 1. \quad (10)$$

Since we assumed $Re(\lambda_2) \geq 0$, there is $Re\left[\frac{1}{(\lambda_2+\mu)}\right] > 0$, and the left-hand side of equation (9) satisfies

$$Re\left(1 + \frac{\mathcal{R}_v E^*(\sigma+\mu)}{S_0} \cdot \frac{1}{\lambda_2+\mu}\right) = 1 + \frac{\mathcal{R}_v E^*(\sigma+\mu)}{S_0} Re\left(\frac{1}{\lambda_2+\mu}\right) > 1.$$

Therefore, taking the norm on the left side of the equal sign in equation (9), we get

$$\left|1 + \frac{\mathcal{R}_v E^*(\sigma+\mu)}{S_0} \cdot \frac{1}{\lambda_2+\mu}\right| > 1. \quad (11)$$

At this point we can see that the inequalities (10) and (11) are contradictory to the equation (9), so the initial assumption about $\lambda_2$ is invalid, that is, all the eigenvalues of the characteristic equation $det(\lambda_2 E - M_2) = 0$ have negative real parts[17]. That is, when $\mathcal{R}_v > 1$, the endemic equilibrium $P^*$ of system (1) is locally asymptotically stable.

**Theorem 6.** When $\mathcal{R}_v > 1$ and $h = \frac{I_1+\varepsilon I_2}{I_1^*+\varepsilon I_2^*} = 1$, the endemic equilibrium $P^*$ of system (1) is globally asymptotically stable.

Obviously, we can get $(1-\theta\omega)\Lambda = \beta S^*(I_1^* + \varepsilon I_2^*) + \mu S^*$, $\mu + \sigma = \frac{\beta S^*(I_1^*+\varepsilon I_2^*)}{E^*}$,

$$\gamma_1 + \mu + \alpha + d_1 = \frac{\rho\sigma E^*}{I_1^*}, \gamma_2 + \mu + d_2 = \frac{(1-\rho)\sigma E^*}{I_2^*} + \frac{\alpha I_1^*}{I_2^*}.$$

Define the same function $h(x)$ as above.

**Proof.** We define a function $V_1$ as follows[20],

$$V_1(\phi) = S^*h\left(\frac{\phi_1}{S^*}\right) + A_1 E^*h\left(\frac{\phi_2}{E^*}\right) + A_2 I_1^*h\left(\frac{\phi_3}{I_1^*}\right) + A_3 I_2^*h\left(\frac{\phi_4}{I_2^*}\right)$$

Where

$$A_1 = \frac{(1-\rho)\sigma E^*}{(\mu+\gamma_2+d_2)I_2^*}, A_2 = \frac{\alpha I_1^*}{\rho\sigma E^*} \cdot \frac{\beta S^*(I_1^*+\varepsilon I_2^*)}{(\mu+\gamma_2+d_2)I_2^*}, A_3 = \frac{\beta S^*(I_1^*+\varepsilon I_2^*)}{(\mu+\gamma_2+d_2)I_2^*}. \quad (12)$$

Define the following expression

$$x = \frac{S}{S^*}, y = \frac{E}{E^*}, z = \frac{I_1}{I_1^*}, w = \frac{I_2}{I_2^*}, h = \frac{I_1+\varepsilon I_2}{I_1^*+\varepsilon I_2^*}.$$

The derivative of $V_1$ with respect to $t(t > 0)$ along $U(t)$ is

$$\frac{dV_1(U(t))}{dt} = -\frac{\mu}{S(t)}[S(t) - S^*]^2 + f(x,y,z,w,h)$$

Where

$f(x,y,z,w,h) =$

$$A_3(1-\rho)\sigma E^*\left[3 + h - \left(\frac{1}{x} + \frac{xh}{y} + w + \frac{y}{w}\right)\right] + A_3\alpha I_1^*\left[3 + h + y - \left(\frac{1}{x} + xh + \frac{y}{z} + w + \frac{z}{w}\right)\right]$$

If $h = 1$, we can get

$f(x,y,z,w,h) =$

$$A_3(1-\rho)\sigma E^*\left[4-(\frac{1}{x}+\frac{x}{y}+w+\frac{y}{w})\right]+A_3\alpha I_1^*\left[4+y-(\frac{1}{x}+x+\frac{y}{z}+w+\frac{z}{w})\right]$$

We know $f(x,y,z,w,h) \leq 0$ and $f(x,y,z,w,h) = 0$ if and only if $x = y = z = w = 1$. There is $\frac{dV_1(t)}{dt} \leq 0$, and $\frac{dV_1(t)}{dt} = 0$ if and only if $U(t) = P^*$. According to the LaSalle's Invariance Principle[19], when $\mathcal{R}_v > 1$ and $h = \frac{I_1+\varepsilon I_2}{I_1^*+\varepsilon I_2^*} = 1$, the the endemic equilibrium $P^*$ of system (1) is globally asymptotically stable.

## 3. Numerical simulation results

### 3.1 Data fitting

To better improve public health, the Chinese government has not only increased public health funding, revised laws on infectious disease control, implemented the world's largest Internet-based disease reporting system, and initiated a program to rebuild local public health facilities [21]. After 2004, China's disease surveillance system began to report detailed data on TB cases. We obtained data for the years 2005-2021 and fitted the TB case data by model (1), and the results of the fit are shown in Figure 5. 5(A) gives the actual case data and the fitted optimal curve, which is in general agreement with the number of reported cases of tuberculosis. 5(B) shows the absolute error between the fitted curve and the real data, where the maximum error value is 75000, and the error value is small compared with the total number of cases. We perform a goodness-of-fit test on the fitting results and the results are shown in Appendix A. The goodness-of-fit coefficient is 0.9546, which is a good fit. The values of the parameters obtained from the fitting are shown in Table 1.

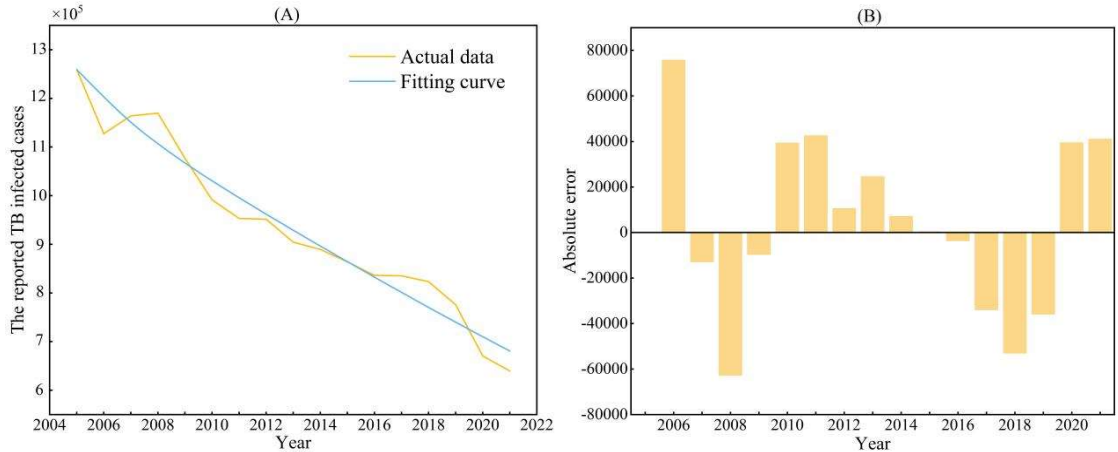

Figure 5 Results of fitting tuberculosis cases, 2005-2021. (A) Real case data and fitted optimal curves. (B) Absolute error between fitted curve and real data.

Table 1. Fitting results of parameters in model (1).

| Parameter | Baseline value | Source | Parameter | Baseline value | Source |
|---|---|---|---|---|---|
| $\Lambda$ | 16245911 | [22] | $\theta$ | 1 | [24] |
| $\mu$ | 0.007 | [22] | $\varepsilon$ | 0.5 | Assumed |
| $\beta$ | $3.35\times10^{-10}$ | Fitted | $\gamma_1$ | 0.005 | Assumed |
| $\sigma$ | 6 | [23] | $\gamma_2$ | 0.496 | [25] |
| $\omega$ | 0.728 | [6] | $d_1$ | 0.06 | [26] |
| $\rho$ | 0.3877 | Fitted | $d_2$ | 0.0025 | [25] |
| $\alpha$ | 0.2514 | Fitted | | | |

We model changes in the number of potentially TB-infected ($I_1$), detected TB-infected ($I_2$), and the sum of this two infected ($I_1+I_2$), and the results are shown in Figure 6. We obtain the mean number of potentially TB-infected patients is 614,186 (95% CI [562,6631,665,741]), and about 39.5% of the patients are potentially infected, see Appendix B for the data; The number of both potentially infected ($I_1$) and detected ($I_2$) tuberculosis infections has been declining over time, with the number of $I_2$ decreasing from 1.25 million in 2005 to 680,000 in 2021, and the number of $I_1$ decreasing from 750,000 in 2005 to 450,000 in 2021. Therefore, to better control the spread of TB, testing efforts need to be increased so that more TB patients can be diagnosed in a timely manner.

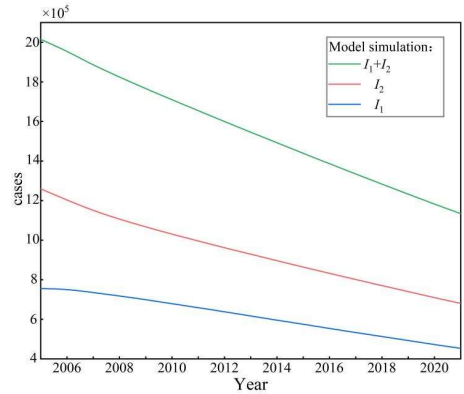

Figure 6. Changes over time in the number of undetected TB $I_1$ and detected $I_2$

### 3.2 Sensitivity analysis

By determining the correlation of each parameter in the model with the basic reproduction number $\mathcal{R}v$, we determine the most sensitive epidemiological parameter in controlling the spread of an epidemic [27]. We perform a sensitivity analysis of the basic reproduction number by means of the partial-rank correlation coefficient (PRCC) [28] in terms of the magnitude of the absolute value of the correlation coefficient, and the results are shown in Figure 7. Parameters lying above the red dashed line indicate a positive correlation with $\mathcal{R}v$ and those below a negative correlation with $\mathcal{R}v$. Parameters in the yellow area have a stronger correlation with $\mathcal{R}v$. The results of the sensitivity analysis show that the annual number of births $\Lambda$, the relative transmission rate $\varepsilon$, the BCG effectiveness rate $\omega$, the BCG vaccination rate $\theta$ are the most sensitive parameters. The proportion of undetected TB infections $\rho$ and the detection rate of undetected TB infections $\alpha$ are strongly correlated, and the results of the correlation coefficients are shown in Appendix C. Simultaneously increasing the detection of potentially infected persons in at-risk populations and improving the recovery rate of persons infected with tuberculosis can significantly reduce the spread of tuberculosis. Vaccination is still the most effective way to reduce the risk of disease transmission.

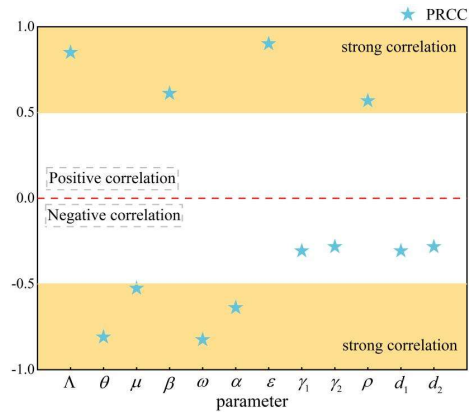

Figure 7. Correlation coefficients between $\mathcal{R}v$ and model parameters (the yellow area is the strong correlation area)

We discuss the relationship between the values of the basic reproduction number $\mathcal{R}_v$ and the values taken by these parameters [29], gave a contour map of $\mathcal{R}_v$, and assess the role of vaccination and detection measures and the results are shown in Figure 8. Figure 8(A) illustrates that the value of the basic reproduction number $\mathcal{R}_v$ decreases rapidly with the gradual increase of the BCG vaccination rate and the BCG effectiveness rate $\omega$, and that there is always $\mathcal{R}_v < 1$ when $\theta\omega > 0.5$. In Fig. 8(B), the value of the basic reproduction number $\mathcal{R}_v$ decreases as the detection rate $\alpha$ increases and the proportion of undetected $\rho$ decreases, and there is always $\mathcal{R}_v < 1$ when $\alpha > 0.4$.

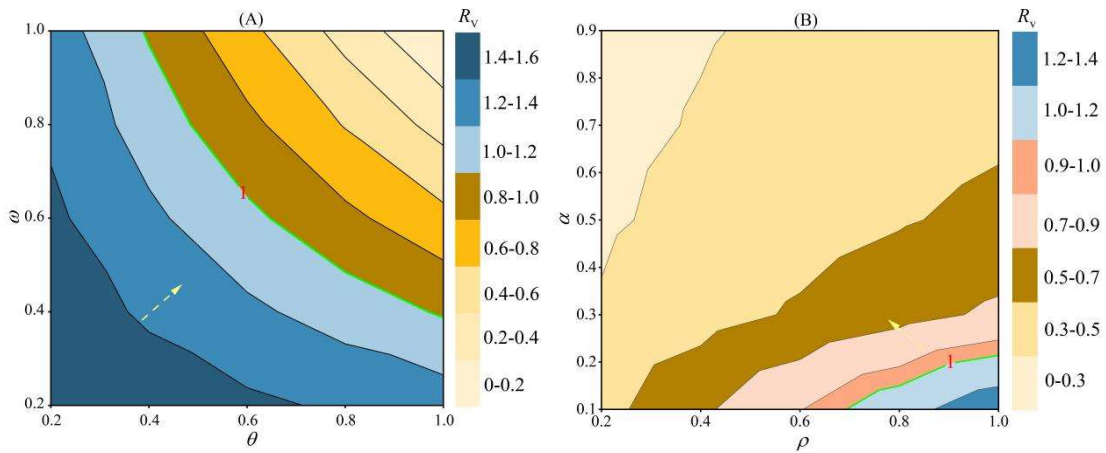

Figure 8. Contour plot of basic reproduction number $\mathcal{R}_v$ versus parameters. (A) $\mathcal{R}_v$ with vaccination rate $\theta$ and effectiveness rate $\omega$. (B) $\mathcal{R}_v$ versus detection rate $\alpha$ and proportion undetected $\rho$. (The green curve is $\mathcal{R}_v = 1$, and the arrow points to the direction where $\mathcal{R}_v$ decreases the fastest.)

**3.3 Impact of detection measures on the spread of the epidemic**

Based on these results, we obtain that vaccination and detection measures can be used to control the spread of tuberculosis. However, in practice, a more direct measure than increasing the vaccination rate and effectiveness of the vaccine is to increase the intensity of detection [17], which is reflected in the model as a decrease in the parameter $\rho$ and an increase in $\alpha$. Therefore, we explore the impact of changes in the parameters $\rho$ and $\alpha$ on the spread of the epidemic. Keeping other parameters constant, we perform simulations for $\rho$ taking values of 20%, 40%, 60% and 80%, and the results are shown in Figure 9. According to Figures 9A, 9B, we obtain that the cumulative number of infected persons rises significantly as the proportion $\rho$ of undetected persons increases. As $\rho$ decreases, the cumulative number of infected persons is lower. Decreasing $\rho$ by 20% reduces the cumulative number of infections by an average of 15 million and the proportion of undetected TB patients by an average of 10.5%, which is a significant decrease. Similarly, we perform simulations for $\alpha$ values of 20%, 40%, 60%, and 80%, and the results are shown in Figures 9C, 9D. We get that as the detection rate $\alpha$ increases, the cumulative number of infected people decreases significantly. When $\alpha$ is larger, the cumulative number of infected persons is lower. When $\alpha$ is increased by 20%, the cumulative number of infections is reduced by an average of 5 million and the proportion of undetected tuberculosis patients is reduced by an average of 7.6%. The simulation results suggest that increased detection intensity is beneficial in controlling the spread of the epidemic.

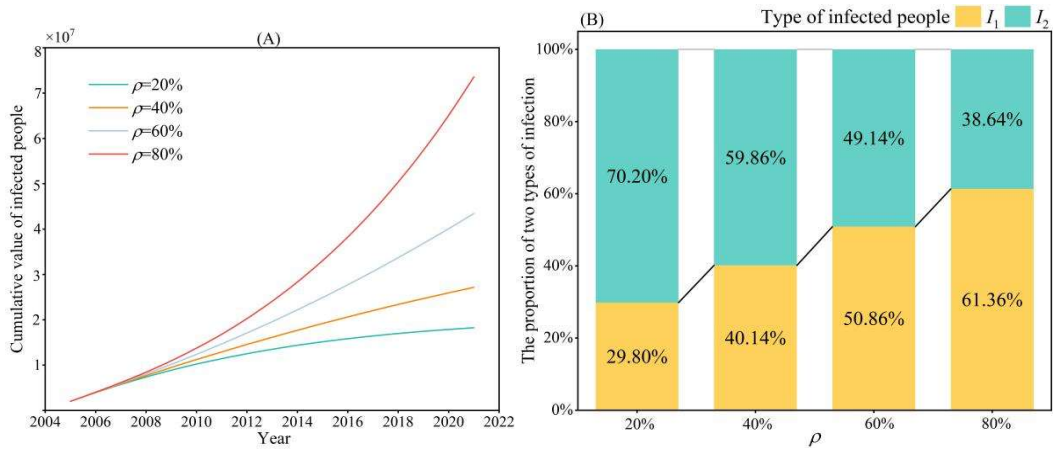

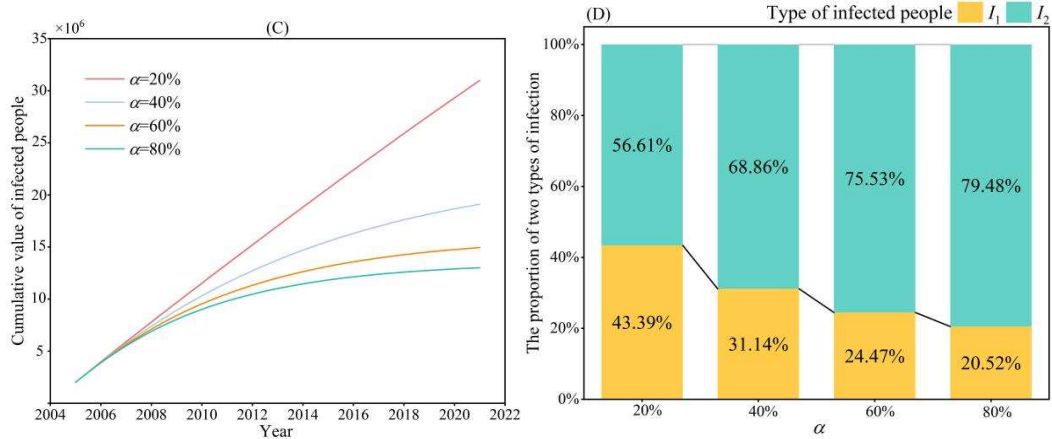

Figure 9. Impact of detection measures on the spread of outbreaks. (A)The effect of changes in the proportion of undetected $\rho$ on the cumulative number of the infected. (B) The effect of changes in the parameter $\rho$ on the proportion of the infected. (C)Effect of changes in detection rate $\alpha$ on the cumulative number of the infected. (D) Effect of the value of the parameter $\alpha$ on the proportion of the infected.

**3.4 Feasibility assessment of reaching WHO End TB Strategy in China**

In controlling the spread of TB, China has taken various measures, a 5-year national program in the 1980s, a 10-year national program to control TB in the 1990s, and the Modern Tuberculosis Control Strategy [30]. With the implementation of these short-term and long-term plans, tuberculosis has been effectively controlled in China. In the nearly 20 years from 2004 to 2021, new cases of TB in China fell from a peak of 1.25 million in 2005 to 630,000 in 2021. However, tuberculosis (TB) remains a huge challenge in China. To end the TB epidemic globally, the World Health Organization (WHO) proposed a post-2015 Global Strategy to End TB in 2014, with the strategic goal of reducing TB incidence by 50% by 2025 (compared to 2015) and reducing new cases by 90% by 2035 [31]. The number of new cases of tuberculosis in China was 864,015 in 2015, and the WHO's goal is for China to reduce the number of new cases of tuberculosis to 86,402 by 2035.

In this section, we assess the feasibility of achieving the goal of ending tuberculosis in China. We examine whether it is feasible to achieve the WHO's End Strategy TB goal based on different control strategies. Using the parameters in Table 1, the basic reproduction number $\mathcal{R}_v = 0.4442$ is calculated. Based on model (1), we simulate the number of new cases of tuberculosis and the results are shown in Figure 10. Our modeling suggests that without additional control measures, there will still be nearly 340,000 new TB infections in China in 2035, failing to meet the WHO's desired target, shown Figure 10(A). We also discuss the impact of different control measures on TB transmission: (1) A population of TB-infected individuals focusing on self-behavioral control and wearing masks voluntarily reduces the rate of transmission $\varepsilon$, and if the rate of transmission can be controlled $\varepsilon < 0.1$, China could reach the expected WHO target by 2035, although this is not easily achievable, shown in Figure 10(B). (2) The government and relevant departments increase the intensity and scope of detection, that is, the proportion of undetected $\rho$ decreases and the rate of detection $\alpha$ increases, when $\alpha = 0.35, \varepsilon = 0.25$ or $\rho = \varepsilon = 0.2$, the number of annual incidence of TB infections in China will decrease rapidly and we can realize the WHO target, shown in Figure 10(C). (3) At the same time, by enhancing the effects of therapeutic drugs and improving the recovery of TB patients, i.e., the recovery rate $\gamma_2$ increases, China can achieve the strategic goal of ending TB, the results are shown in Figure 10(D). In summary, a combination of control measures to reduce transmission rates, increase testing intensity, and improve recovery rates can reduce the number of TB cases to low levels in a short period of time.

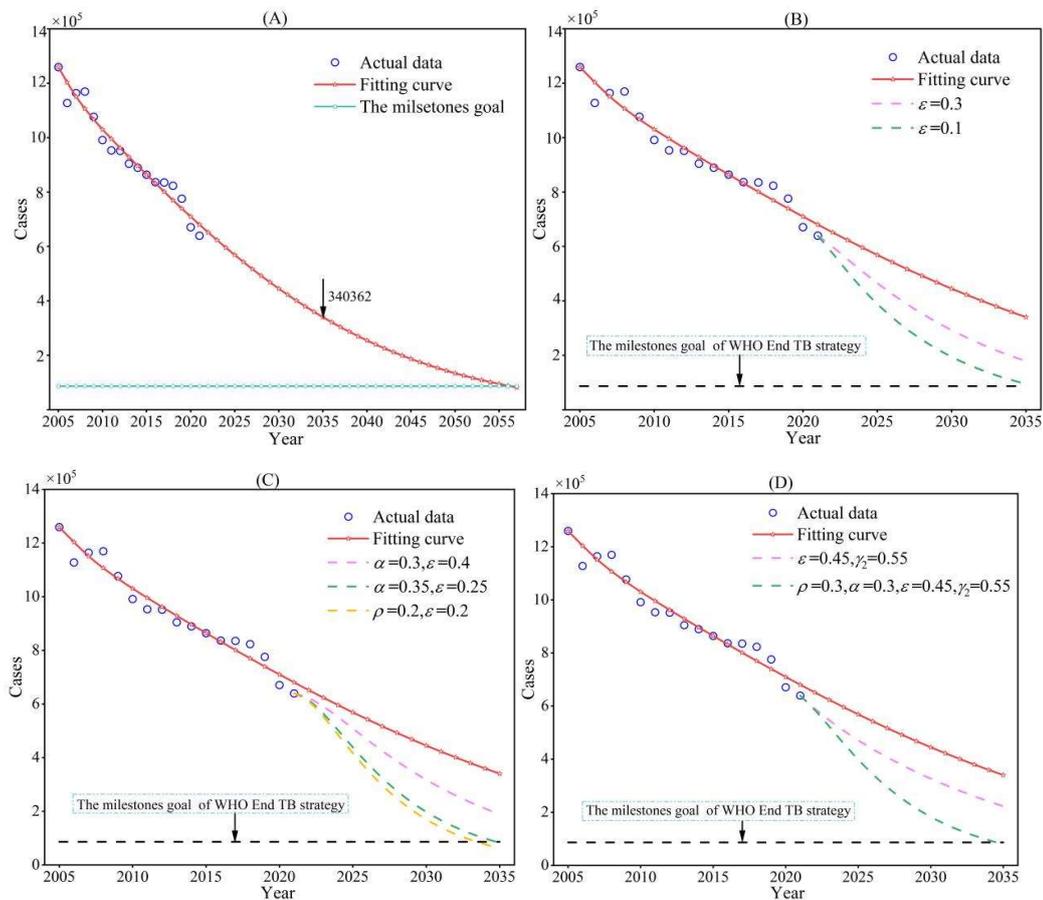

Figure 10. Results of the feasibility assessment of achieving the strategic goal of ending tuberculosis in China. (A) Annual incidence of tuberculosis in China under current measures. (B) Control of transmission rate $\varepsilon$ of the annual incidence of simulation prediction. (C) Simulated

prediction of annual incidence for increased detection intensity and detection coverage. (D) Simulated prediction of annual incidence rates for improved recovery.

**4. Discussion**

China has one of the highest incidence rates of tuberculosis in the world. Due to the relatively complex course of human infection with Mycobacterium tuberculosis, the hidden nature of tuberculosis and the long treatment period, there are still potential tuberculosis infected people in the population, which poses a lot of challenges for the prevention and control of tuberculosis. Accurate assessment of the number of potentially tuberculosis-infected individuals and selection of feasible control strategies are essential for tuberculosis prevention and control.

We collect data related to tuberculosis in China from 2004 to 2021 and analyze the trend of the number of tuberculosis cases in China. The annual number of new cases of tuberculosis in China has been declining in recent years, and the decline has been more rapid in the eastern and southern regions; however, the number of new cases of tuberculosis in China remains high. In this study, a tuberculosis transmission dynamics model including the potentially infected and vaccination is developed to estimate the number of potentially infected persons using data related to tuberculosis reported in China, and the results show that the number of potentially infected persons with tuberculosis in China is about 610,000, which accounted for 39.5% of the total number of tuberculosis cases. This finding is consistent with the results obtained from the Chinese evidence-based guidelines for community-based active screening for tuberculosis [32] based on the research data, which indicates that our findings are reliable, and that it is feasible to assess the number of potentially infected individuals in the population based on the number of existing cases and the establishment of a model of infectious disease dynamics based on the mechanism of transmission, which is one of the greatest innovations of this paper. Without additional control measures, there will still be nearly 340,000 new tuberculosis infections in China in 2035, making it impossible to achieve the strategic goal of ending tuberculosis. A combination of control measures, such as increased testing and improved drug treatment, is optimal if the number of tuberculosis cases is to be reduced to a low level in a short period of time. To eliminate tuberculosis, China needs to continue to strengthen the implementation of tuberculosis control measures and further explore other effective control measures.

Limitations of this paper, we estimate the number of potentially TB-infected individuals using national TB incidence data, but the model omit the length of effective protection from BCG vaccine or reinfection of recovered individuals. In a follow-up study, we will take these factors into account for a more in-depth study of the model.

**Acknowledgement:** This study was funded by the Guangdong Provincial Key Laboratory of Functional Substances in Medicinal Edible Resources and Healthcare Products (2021B1212040015）and Natural Science Foundation of Shandong Province (Grant No.

ZR2023QA059), the Pyramid Talent Training Project of BUCEA (JDYC20200327). We thank all the individuals who generously shared their time and materials for this study.

**Appendix A. Calculation of the goodness-of-fit coefficient**

We use $R^2$ to denote the goodness-of-fit coefficient, and $R^2$ is defined by the expression (13).

$$R^2 = 1 - \frac{RSS}{TSS} \tag{13}$$

Where, *RSS* means residual sum of squares, which represents the sum of squares of the deviations between the actual and simulated values. *TSS* means total sum of squares, which represents the sum of squares of the deviations between the actual and expected value[33].

We use the magnitude of the goodness-of-fit coefficient to judge the effectiveness of the fit. The closer the goodness-of-fit coefficient is to 1, the better the fit is.

**Appendix B. Additional notes to Figure 6**

Table 2. Simulation results of the model (Supplementary Figure 6, 其中 NTB 表示未被发现的肺结核感染者的数量,TB 表示已发现的肺结核感染者的数量,PTB 表示所有的肺结核感染者中 NTB 的比例).

| | NTB | TB | PTB | | NTB | TB | PTB |
|---|---|---|---|---|---|---|---|
| Year | $I_1$ | $I_2$ | $I_1/(I_1+I_2)$ | Year | $I_1$ | $I_2$ | $I_1/(I_1+I_2)$ |
| 2005 | 755584 | 1259308 | 37.50% | 2014 | 595872 | 896439 | 39.93% |
| 2006 | 750370 | 1203312 | 38.41% | 2015 | 574901 | 864319 | 39.95% |
| 2007 | 735406 | 1151093 | 38.98% | 2016 | 554042 | 832592 | 39.96% |
| 2008 | 718005 | 1106785 | 39.35% | 2017 | 533363 | 801261 | 39.96% |
| 2009 | 699074 | 1067254 | 39.58% | 2018 | 512917 | 770357 | 39.97% |
| 2010 | 679198 | 1030593 | 39.72% | 2019 | 492736 | 739905 | 39.97% |
| 2011 | 658698 | 995793 | 39.81% | 2020 | 472869 | 709955 | 39.98% |
| 2012 | 637885 | 962023 | 39.87% | 2021 | 453344 | 680544 | 39.98% |
| 2013 | 616902 | 928975 | 39.91% | | | | |

**Appendix C. Additional notes to Figure 7**

Table 3. The correlation coefficient between the basic reproduction number and each parameter of the model. (Supplementary Figure 7).

| Parameter | PRCC | Parameter | PRCC | Parameter | PRCC |
|---|---|---|---|---|---|
| $\Lambda$ | 0.850 | $\omega$ | -0.825 | $\gamma_2$ | -0.283 |
| $\theta$ | -0.810 | $\alpha$ | -0.638 | $\rho$ | 0.568 |
| $\mu$ | -0.525 | $\varepsilon$ | 0.901 | $d_1$ | -0.307 |
| $\beta$ | 0.611 | $\gamma_1$ | -0.307 | $d_2$ | -0.283 |